\documentclass[sigconf]{acmart}
\AtBeginDocument{%
  }

\usepackage{multirow}
\usepackage{hyperref}
\usepackage{enumitem}

\graphicspath{{./fig/}}

\copyrightyear{2024}
\acmYear{2024}
\setcopyright{acmlicensed}\acmConference[CHI '24]{Proceedings of the CHI Conference on Human Factors in Computing Systems}{May 11--16, 2024}{Honolulu, HI, USA}
\acmBooktitle{Proceedings of the CHI Conference on Human Factors in Computing Systems (CHI '24), May 11--16, 2024, Honolulu, HI, USA}
\acmDOI{10.1145/3613904.3641906}
\acmISBN{979-8-4007-0330-0/24/05}

\begin{document}

\title{Examining the Role of Peer Acknowledgements on Social Annotations: Unraveling the Psychological Underpinnings}
\acmConference[CHI'24]{May.11-16}{May.11-16}{Honolulu, HI,
USA}

\author{Xiaoshan Huang}
\email{xiaoshan.huang@mail.mcgill.ca}
\affiliation{%
  \institution{Department of Educational \& Counselling Psychology\\ McGill University}
  \city{Montreal}
  \country{Canada}
}

\author{Haolun Wu}
\email{haolun.wu@mail.mcgill.ca}
\affiliation{%
  \institution{School of Computer Science\\ McGill University}
  \city{Montreal}
  \country{Canada}
}

\author{Xue Liu}
\email{xueliu@cs.mcgill.ca}
\affiliation{%
  \institution{School of Computer Science\\ McGill University}
  \city{Montreal}
  \country{Canada}
}

\author{Susanne Lajoie}
\email{susanne.lajoie@mcgill.ca}
\affiliation{%
  \institution{Department of Educational \& Counselling Psychology\\ McGill University}
  \city{Montreal}
  \country{Canada}
}


\renewcommand{\shortauthors}{Huang et al.}

\newcommand{\HL}[1]{{\color{red} [Haolun: ``#1'']}}
\newcommand{\new}[1]{\textcolor{black}{#1}}

\begin{abstract}
This study explores the impact of peer acknowledgement on learner engagement and implicit psychological attributes in written annotations on an online social reading platform. Participants included 91 undergraduates from a large North American University. Using log file data, we analyzed the relationship between learners’ received peer acknowledgement and their subsequent annotation behaviours using cross-lag regression. Higher peer acknowledgements correlate with increased initiation of annotations and responses to peer annotations. By applying text mining techniques and calculating Shapley values to analyze 1,969 social annotation entries, we identified prominent psychological themes within three dimensions (i.e., affect, cognition, and motivation) that foster peer acknowledgment in digital social annotation. These themes include positive affect, openness to learning and discussion, and expression of motivation. The findings assist educators in improving online learning communities and provide guidance to technology developers in designing effective prompts, drawing from both implicit psychological cues and explicit learning behaviours.
\end{abstract}

\ccsdesc[500]{Applied computing~Learning management systems}
\ccsdesc[500]{Applied computing~Collaborative learning}
\ccsdesc[500]{Human-centered computing~Empirical studies in HCI}
\ccsdesc[500]{Computing methodologies~Feature selection}
\ccsdesc[500]{Computing methodologies~Information extraction}

\keywords{Digital social annotation, peer acknowledgment, Shapley value, text mining, learner behavior, psychological themes}

\maketitle

\section{Introduction}
Digital social annotation, a platform where readers comment within text margins while reading, has evolved into a dynamic arena for computer-supported collaborative learning (CSCL) ~\cite{krouska2018social}, facilitating peer interactions ~\cite{adams2020building} and co-construction of knowledge ~\cite{morales2022using}. 
However, the quality of online discourse varies, affecting learners' interactive experience and learning outcomes ~\cite{garrison2005facilitating}. 
Peer acknowledgement, often conveyed through positive emoticons or supportive words from fellow users, plays a pivotal role in shaping user behaviour, both in social media ~\cite{ge2018emoji,lee2017making} and online learning environments ~\cite{phirangee2016loving}. 
However, despite extensive systematic reviews on social annotation (e.g., ~\cite{zhu2020reading, ghadirian2018social}), limited attention has been devoted to understanding how peer acknowledgement influences learners' subsequent behaviours within social annotation platforms. 
Moreover, there remains a gap in our understanding of the key factors that establish evaluative criteria for the quality of written content and how this written content influences peers' reactions. 
Consequently, it is imperative to delve into the examination of linguistic features within written annotations, as they provide valuable insights into the psychological aspects of CSCL. 
An in-depth exploration of the interplay between these dimensions can inform the development of effective instructional strategies, fostering student engagement with course materials and within the online learning community ~\cite{huang2023social}.
This study aims to address the existing gap by exploring the impact of peer acknowledgement on learners' subsequent annotation behaviours and comparing the relative importance of linguistic indicators within four psychological dimensions (i.e., affect, cognition, motivation, and social) in written annotations related to peer acknowledgement.
Given that the digital social annotation platform provides rich trace data without interrupting learners' natural learning processes, it is recommended to apply learning analytics to examine their ``actual'' learning behaviours ~\cite{jovanovic2017learning} within this context.
The following sections will delve into the theoretical foundations, draw from previous empirical studies, and detail the human-centred analytical methods employed in this study.

\section{Background}
\subsection{Peer Acknowledgement in Digital Social Annotation} 
Digital social annotation is a productive forum for online learning ~\cite{morales2022using, licastro2019past}. 
It employs a computer-supported communication approach to facilitate interactions among peers within a shared community or individuals with similar interests in specific topics. More precisely, users can highlight valuable information by adding annotations to selected text and respond to others' annotations through written text or by using emoticons to express reactions. 
This functionality facilitates the co-construction of knowledge ~\cite{morales2022using}, nurtures a sense of collaborative learning community ~\cite{johnson2010individual}, and transforms individual learning styles to social ones by promoting flexible asynchronous conversations ~\cite{blyth2014exploring}. 
In the digital era, social annotation has emerged as a prominent pedagogical approach in higher education, enhancing online interaction without impeding collaborative learning among students. However, the quantitative and qualitative aspects of online discourse can vary, in terms of engagement in annotation activities and the content contributed, respectively.

Furthermore, the quality of interaction, especially as indicated by the responses and reactions of peers, significantly influences online users' behaviours. Individuals may experience feelings of isolation if they perceive a lack of connection with others in virtual spaces ~\cite{huangemotion}. 
In contrast, the perceived richness of online discussion forums has a significant positive effect on student
participation, interaction, and learning ~\cite{balaji2010student}.
For example, a previous study of a social media platform known for knowledge sharing, identified peer acknowledgement as a significant indicator of the recipient's comment behaviour, positively predicting more frequent and longer comment posting ~\cite{burtch2022peer}. 
A recent systematic review of empirical studies in social annotation has found that the majority of previous research has focused on assessing learners' performance by quantifying annotation metrics ~\cite{ghadirian2018social}. 
However, despite these findings, limited research has explored how peer acknowledgement, in the form of receiving ``upvotes'' emoticons, impacts learners' subsequent behaviour in digital social annotation contexts. 
This gap in the literature underscores the need to investigate the influence of peer acknowledgement on learners' behaviours and the linguistic factors that affect peer reactions in such environments.

\subsection{Revealing Psychological Signs through Linguistic Features in Computer-Supported Collaborative Learning}  
Social annotations can significantly benefit students' learning when used effectively~\cite{bradley2007supporting}. 
There is compelling evidence supporting the positive impact of annotations on students' memory and learning (e.g., ~\cite{shadiev2015impact, van2005research}). 
Moreover, students who exhibit higher levels of critical thinking and connect their personal reflections to the course materials tend to achieve superior outcomes compared to peers who simply replicate provided content or engage in casual conversations to demonstrate social skills ~\cite{wickersham2006content}. However, some learners may not be skilled enough to identify the key concept of a given content ~\cite{johnson2010individual}, which may lead to reflecting at a surface level. On another occasion, learners with lower working memory capacity may shift their focus towards more superficial conversation when tackling multiple sources ~\cite{plass2003cognitive}. In an autonomous context such as digital social annotation, interactive actions among learners may reflect the quality of a written text;
\new{there is an increasing demand to employ computer processing of discourse to highlight crucial exchanges for the benefit of teachers and other stakeholders ~\cite{siqin2015fixed,stahl2015decade}.}
Thus, cultivating the written text of annotation will give a fuller picture of what makes a high-quality annotation, as observed and acknowledged by peers.

The investigation of linguistic features becomes even more critical when considering the four dimensions of psychological factors: affect, cognition, motivation, and social. 
According to the community of inquiry framework~\cite{garrison2007researching}, learners' online learning experience is influenced by their perceived emotional feelings, interactions and cohesiveness within the community, which are highly correlated to their cognitive procedures ~\cite{huang2023social}. Studies by Pekrun and colleagues ~\cite{pekrun2002positive} emphasize the role of emotions in learning, highlighting how emotions can significantly impact students' comprehension and retention of information. 
Additionally, the cognitive dimension encompasses various aspects of memory, cognitive processes, and information processing ~\cite{anderson1983neuromathematical}, all of which are deeply intertwined with the way students engage with and make use of annotations and comments in their learning process. 
Furthermore, Deci and Ryan ~\cite{ryan2000self} have extensively explored the motivational aspect, emphasizing the importance of intrinsic motivation in driving effective learning. 
Finally, the socio-cultural theory ~\cite{vygotsky1978} underscores the influence of social interactions on cognitive development and learning. 
It emphasizes the significance of collaborative and social learning environments.

Therefore, the investigation into linguistic features, within the social annotation context, allows us to gain a comprehensive understanding of how these psychological dimensions influence students' learning experiences. 
This understanding can inform instructional approaches that promote effective learning strategies and foster a deeper engagement with course materials and within the learning community. 

\subsection{Aim of This Study} 
In this study, we aim to answer the following two research questions (RQs):

\textit{\textbf{RQ1.} How does peer acknowledgement influence learners’ social annotation behaviours?} 

\textit{\textbf{RQ2.} What are the predominant linguistic features that affect peer acknowledgement?}

We investigate the two research questions through a series of learning analytics. 
We hypothesize that peer acknowledgement, indicated by more upvotes received at time $t$-1, leads to more initiated annotations and responses posted at time $t$, exceeding the average of peers' contributions.
Additionally, we aim to examine the impact of learners' psychological factors on peer acknowledgement by employing text-mining and Shapley value techniques.
These methods help us identify the most significant indicators for peer acknowledgement by comparing emotional, cognitive, motivational, and social perspectives. 
We will discuss the identified influential predictive factors from these four dimensions in detail.

\section{Methodology}

\subsection{Participants}
This study involved 91 undergraduate students (53 females and 38 males) who were enrolled in the course ``Integrating Educational Technology in Classrooms'' at a large North American university in Fall 2021. 
The average age of the participants was 20.67 ($SD$ = 1.25). 
The course was provided by the Faculty of Education and was open to undergraduate students from various academic disciplines, such as Medicine, Arts, Management, and Engineering. 
Therefore, the participants represent students from diverse academic backgrounds.
Over the semester of twelve consecutive weeks, the students were assigned to complete collaborative reading activities using an asynchronous platform called \textit{Perusall}\footnote{\href{https://www.perusall.com/}{https://www.perusall.com/}}. 
The study obtained ethical approval from the university's Research Ethics Board Office.

\subsection{Learning Environment, Procedure, and Data Sources}
\begin{figure*}
    \centering
    \includegraphics[width=1.0\linewidth]{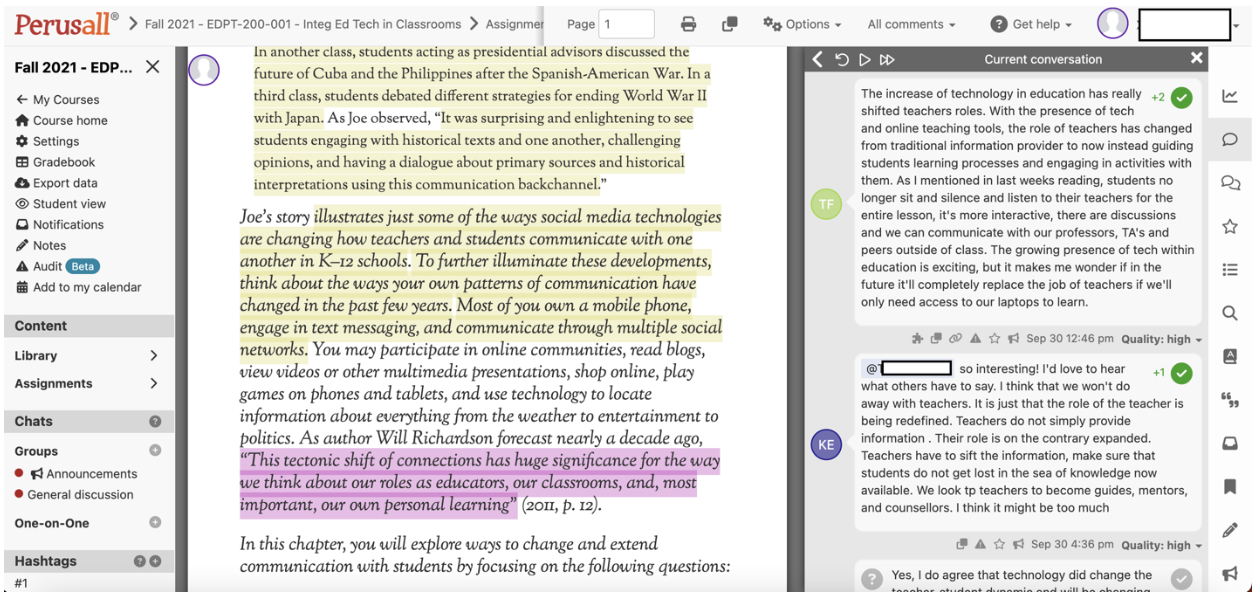}
    \caption{The Interface of the Social Annotation Platform Perusall.}
    \label{fig:Perusall}
\end{figure*}

We use \textit{Perusall} as the major learning context in this study. \textit{Perusall} is an interactive and collaborative annotation platform that empowers educators to design collaborative reading sessions utilizing digital textbooks, documents, and online sources. 
It redefines solitary reading assignments into collaborative social and educational exercises~\cite{adams2020building}. 
During the semester, students were directed to annotate class readings weekly through the acts of sharing comments or responses. These annotations created by students were made visible to their classmates (see Figure~\ref{fig:Perusall}). 
Students could initiate an annotation by leaving a comment under the annotated content. 
Additionally, students could respond to annotations generated by their peers. Furthermore, while this participation was not mandatory and would not contribute to the course credit, students could express their support to peers' annotation using the ``upvote'' feature. 

The assigned reading materials on \textit{Perusall} encompassed 29 chapters from books and research articles, including topics related to theories and practical applications of educational technologies. These assigned readings covered a broad spectrum, including comprehending challenges in educational technology, fostering virtual learning through online resources, and engaging in collaborative activities with social technologies.  

The data sources analyzed in this study were exclusively retrieved from \textit{Perusall}'s log files. 
To achieve the research purpose of this study, we specially examined the extracted interactive behaviors of learners using three data sources:  (\romannumeral1) learners’ interactive behaviors (i.e., number of initiated annotations and responses posted), (\romannumeral2) peer acknowledgement indicated by received peer upvotes, and (\romannumeral3) their written content on each annotation. 

\subsection{Analytic Methods}

We leveraged a range of data mining techniques to address our research questions. We used Cross-lag regression~\cite{kenny1975cross} to delve into the role of peer acknowledgement in steering learners' future social annotation activities in a sequential time context. Furthermore, we adopted text mining strategies to pull linguistic features from annotations, encapsulating affect, cognition, motivation, and social facets.
To pinpoint the crucial features influencing peer acknowledgment in annotations, we applied the Shapley value technique~\cite{roth1988shapley}. Subsequent sections elaborate on these methods.

\paragraph{Cross-lag regression}
To study the lagged influence of peer acknowledgement on learner annotation behaviors, we employed the Cross-lag regression analysis~\cite{kenny1975cross}. 
This approach is particularly useful in examining the lagged effects of one variable on another over distinct time intervals, and is widely used in various domains including education and psychotherapy~\cite{pianta2014cross, notsu2023enhancing}. 
In our context, it helps unravel the complex relationship between peer acknowledgment and the frequency of students initiating annotations and responding to their peers' notes. 
The Cross-lag regression can be expressed as: 
\begin{align}
    Y_t = \beta_0 + \beta_t\cdot X_{t-1} + \epsilon_t,
\end{align}
where \(Y_t\) is the annotation behavior at time \(t\); \(X_{t-1}\) is the peer acknowledgment at the previous time step; \(\beta_t\) denotes the cross-lagged effect; and \(\epsilon_t\) represents the residual error at time \(t\). The intercept, \(\beta_0\), signifies the initial state of annotation behavior unaffected by peer acknowledgment. 
\new{Specifically, time $t$ and $t$-1 are defined based on the order of 29 assigned readings distributed over 12 weeks. Each week involves the assignment of one to three readings for annotation activities, depending on the overall workload for that specific week.}
Given the asynchronous nature of social annotation activities, where students often complete annotations before receiving feedback, understanding this lagged relationship becomes vital. Utilizing this method, we intend to uncover the subtle influences of lagged peer acknowledgment on ensuing reading behaviors in social annotation settings.




\paragraph{Text Mining} 

To address our second research question, we applied text mining techniques to extract linguistic features from student comments using the Linguistic Inquiry and Word Count (LIWC) tool\footnote{\href{https://www.liwc.app/}{https://www.liwc.app/}} developed by Boyd et al.~\cite{boyd2022development}. 
This tool applies natural language processing (NLP) to quantify words and phrases corresponding to specific psychological dimensions~\cite{tausczik2010psychological}, allowing us to explore the psychological attributes reflected in the students' annotations. 
\new{It categorizes words into different linguistic and psychological dimensions covering emotions, cognitive processes, social aspects, and linguistic style. Utilizing LIWC for linguistic feature extraction involves the following five steps: text input, tokenization, lookup in the LIWC dictionary, category counting, and analysis ~\cite{pennebaker2001linguistic}.}

LIWC has facilitated research in various fields, aiding in the analysis of emotional expression and cognitive processes in learning~\cite{huangemotion, zhengthinking, lidetection}. 
Using LIWC in our study reveals the nuanced aspects of student interactions in social annotations, enhancing our understanding of affective, cognitive, motivational, and social dynamics.

\paragraph{Shapley Value}
To investigate and filter those linguistic features that have the highest effect on peer acknowledgment, one effective approach is utilizing the Shapley value technique~\cite{roth1988shapley} — a concept borrowed from cooperative game theory~\cite{fudenberg1991game}. 
The Shapley value assists in distributing the total effect fairly over various contributing factors such as the quality of teaching materials, instructor’s teaching style, course difficulty, among others that are present in the diverse features influencing the feedback.

Formally, the Shapley value $\Phi_i$ of a player $i$ (in our context, a \textit{feature}) in a cooperative game is defined as:
\begin{align}
\Phi_i = \sum_{S\subseteq N \setminus \{i\}} \frac{|S|!(|N|-|S|-1)!}{|N|!} \Big( M(S \cup \{i\}) - M(S) \Big),
\label{eq:shapley}
\end{align}
where the descriptions of notations are as follows:
\begin{itemize}
    \item $N$ is the set of all players.
    \item $S$ is a subset of players not including player $i$.
    \item $M(S)$ is the characteristic function/model that outputs a value to each input coalition $S$.
    \item $|S|$ and $|N|$ denote the cardinality of set $S$ and $N$ respectively.
\end{itemize}

By evaluating all possible subsets of the linguistic features and analyzing the marginal contribution of each feature within these subsets, we obtain a detailed understanding of the individual impact of each feature on peer acknowledgement, as indicated by the score of upvotes received.
The Shapley values derived from this analysis not only quantify the importance of each feature but also facilitate a hierarchical comprehension of their respective influences, ultimately guiding focus towards the most critical aspects to enhance peer acknowledgement.

\section{Result} 
\subsection{RQ1. How does peer acknowledgement influence learners’ social annotation behaviors? }


To investigate our first research question, we carried out two cross-lag regression analyses focusing on the interplay between peer acknowledgment and the distinct dependent variables, namely, the number of annotations initiated and responses given, respectively. 
Before these analyses, we performed preliminary data cleaning. 
The anticipated 2,639 data entries — calculated from 91 participants and 29 reading materials — was reduced to 2,205 due to incomplete inputs from 47 participants. Following z-score normalization~\cite{devore2015probability} and outlier exclusion, we finalized 1,989 data entries deemed relatively independent given the diverse reading materials and one-semester time frame.

Our analysis accounts for potential individual-specific effects by implementing a linear mixed-effects approach. 
\new{To address individual variability within the model, we specifically incorporated a random slope for the grouping variable (i.e., ID), which allows for individual variability in the effect of the predictor variable ~\cite{brauer2018linear}. In our study, the random slope linear mixed model captures individual differences in how the number of peer acknowledgements affect their annotation behaviours. 
We show the details in Appendix~\ref{A}.}

The result evidenced a significant relationship between received peer acknowledgment at time $t$-1 and initiated annotation behaviors at time $t$, with a substantial coefficient of 0.187 ($p$<.001). 
This confirms our hypothesis that increased peer acknowledgment positively correlates with a surge in subsequent annotation activities. 
This result underscores the predictive power of lagged peer acknowledgment in determining future annotation behaviors, sustaining a prominent role in fostering engagement in social reading activities.
Similarly, we explored the impact of lagged peer acknowledgment on the frequency of response in subsequent social annotation activities, recording a significant coefficient of 0.052 ($p$=.009). 
This aligns with our preliminary hypothesis, asserting a positive predictive relationship between peer acknowledgment at time $t$-1 and the learners' responsive engagement at time $t$.
\new{Furthermore, the reciprocal relationship between each annotation behavior (i.e., numbers of initiated annotations and responses given) at $t$-1 and peer acknowledgement at $t$ is non-significant, indicating that the previous annotation behaviors by students do not necessarily influence peer acknowledgment in the next annotation activities.}
 


\subsection{RQ2. What are the predominant linguistic features that affect peer acknowledgement?} 

In answering the second question, we conducted a text-mining analysis of learners’ written comments. 
Specifically, we selected twenty linguistic features from the LIWC dictionary covering four dimensions that match best with our learning context, including affect, cognition, motivation, and social (see Table~\ref{tab:text_mining_analysis}). 
\new{The mean value represents the average linguistic features associated with each selected indicator across all annotation entries, as determined by the LIWC analysis. A higher mean score indicates a greater percentage of words in the annotations expressing the selected indicator.}
Each dimension contains features (indicators) that can be categorized in that field. 
For example, written words that reflect a degree of bravado in the sense of certainty such as ``\textit{really}'', ``\textit{actually}'', and ``\textit{of course}'' are categorized into the cognition dimension, while words that reflect motives in higher achievement (e.g., better) are categorized into the motivation dimension~\cite{boyd2022development}.

\begin{table*}[t]
\caption{Summary and text-mining analysis of twenty linguistic features (indicators) from the LIWC dictionary.}
\resizebox{1.0\textwidth}{!}{
\centering
\begin{tabular}{l|l|l|l|c|c}
\toprule
\textbf{Dimension} & \textbf{Indicator} & \textbf{Description} & \textbf{Sample words} & \textbf{Mean score} & \textbf{Std. deviation} \\
\midrule
\multirow{5}{*}{Affect} & \textit{curiosity} & Expression of curiosity & wonder, look for, research & 0.97 & 1.02 \\
\cline{2-6}
& \textit{emo\_pos} & Expression of positive emotion & happy, hope, good & 0.62 & 0.76 \\
\cline{2-6}
& \textit{emo\_anx} & Expression of anxiety & worry, nervous, afraid & 0.07 & 0.20 \\
\cline{2-6}
& \textit{tone\_pos} & Positive tone & support, well, new & 3.31 & 1.77 \\
\cline{2-6}
& \textit{tone\_neg} & Negative tone & bad, wrong, too much & 0.82 & 0.90 \\
\midrule
\multirow{5}{*}{Cognition} & \textit{cause} & Causation & make, because, why & 2.94 & 1.54 \\
\cline{2-6}
& \textit{certitude} & Boasting of certainty & really, actually, of course & 1.05 & 0.96 \\
\cline{2-6}
& \textit{differ} & Differentiation & but, not, if & 4.01 & 1.71 \\
\cline{2-6}
& \textit{insight} & Insight & know, think, make & 5.30 & 2.21 \\
\cline{2-6}
& \textit{tentat} & Tentative & something, or, any & 2.59 & 1.52 \\
\midrule
\multirow{5}{*}{Motivation} & \textit{achieve} & Achievement & work, better, best & 1.90 & 1.38 \\
\cline{2-6}
& \textit{acquire} & Acquire & get, take, getting & 0.59 & 0.67 \\
\cline{2-6}
& \textit{affiliation} & Affiliation & our, us, help & 1.78 & 1.55 \\
\cline{2-6}
& \textit{power} & Words with power ideas & order, allow, own & 0.85 & 0.83 \\
\cline{2-6}
& \textit{want} & Want & hope, want, wish & 0.19 & 0.35 \\
\midrule
\multirow{5}{*}{Social} & \textit{assent} & Assent & yeah, okay, yes & 0.55 & 0.71 \\
\cline{2-6}
& \textit{comm} & Acts of communication & tell, thank, said & 1.51 & 1.30 \\
\cline{2-6}
& \textit{prosocial} & Prosocial behavior & care, help, please & 0.60 & 0.76 \\
\cline{2-6}
& \textit{socbehav} & Social behavior & say, said, love & 4.46 & 1.99 \\
\cline{2-6}
& \textit{socrefs} & Social referents & you, we, he/she & 6.67 & 2.86 \\
\bottomrule
\end{tabular}}
\label{tab:text_mining_analysis}
\end{table*}

\begin{table*}[t]
\caption{Evaluation and comparison of different machine learning models. The R-squared (R$^2$) is the higher the better, while the Mean Squared Error (MSE) is the lower the better. The best model is bolded.}
    \centering
    \begin{tabular}{l|c|c|c|c}
    \toprule
    \multirow{2}{*}{Model} & \multicolumn{2}{c|}{cross-validation} & \multicolumn{2}{c}{testing}\\
     & R-squared (R²) & Mean Squared Error (MSE) & R-squared (R²) & Mean Squared Error (MSE)\\
    \midrule
    Random Forest & 0.8794 & 0.2531 & 0.8682 & 0.2621 \\
    Lasso & 0.8217 & 0.3132 & 0.8092 & 0.3207 \\
    KNN & 0.7911 & 0.3512 & 0.7794 & 0.3603 \\
    ElasticNet & 0.8123 & 0.3305 & 0.8004 & 0.3410 \\
    \textbf{GBR} & \textbf{0.9231} & \textbf{0.1490} & \textbf{0.9123} & \textbf{0.1561} \\
    \bottomrule
    \end{tabular}
    \label{tab:result_table}
\end{table*}

\begin{figure}[t]
    \centering
    \includegraphics[width=1\linewidth]{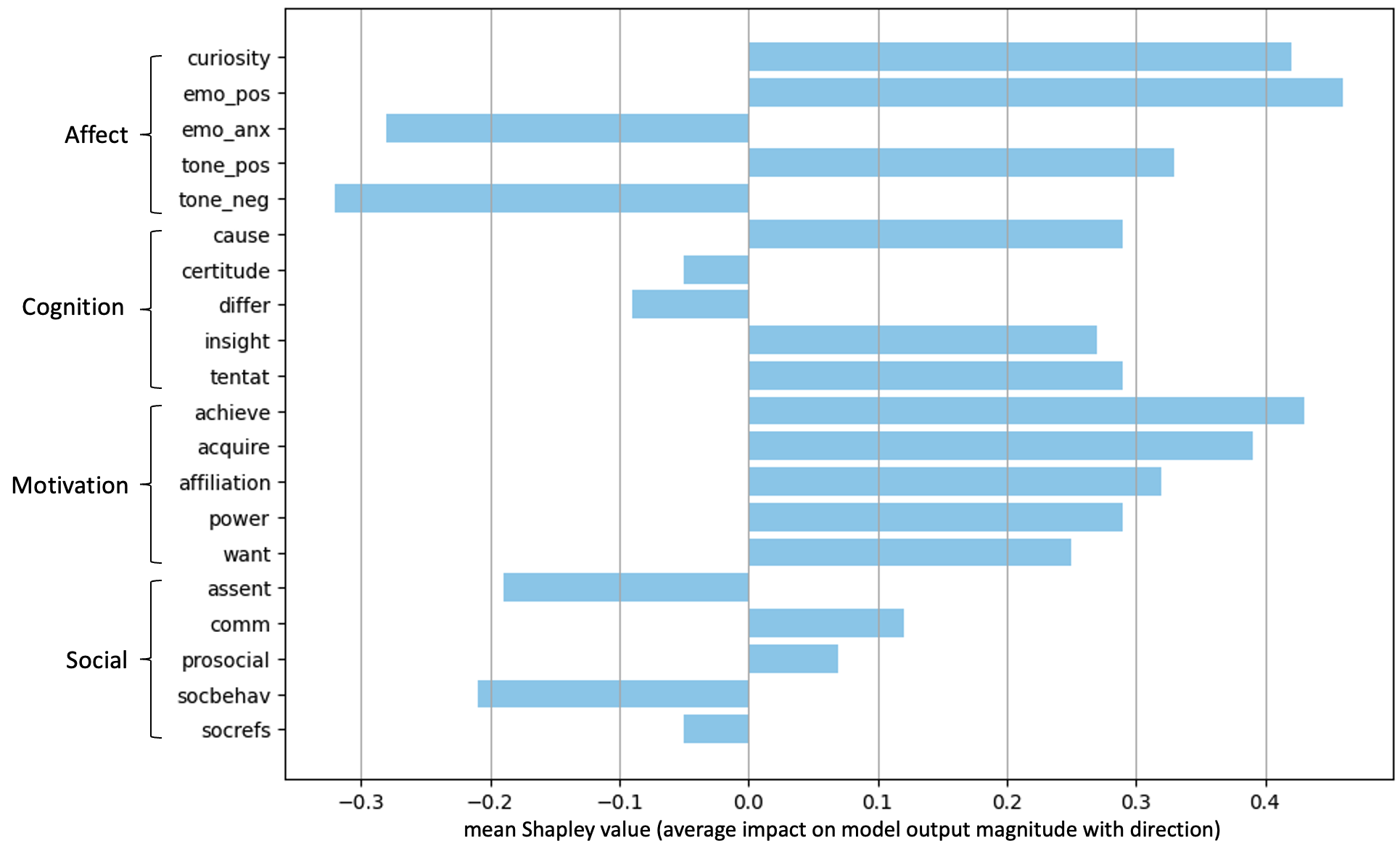}
    \caption{Summary of the Shapley values across all 20 linguistic features, categorized into four dimensions. Each value signifies the average influence of its respective feature on the model’s output, illustrating not only the magnitude but also the direction of the impact.}
    \label{fig:Feature_importance}
\end{figure}

\begin{figure}[t]
    \centering
    \includegraphics[width=1\linewidth]{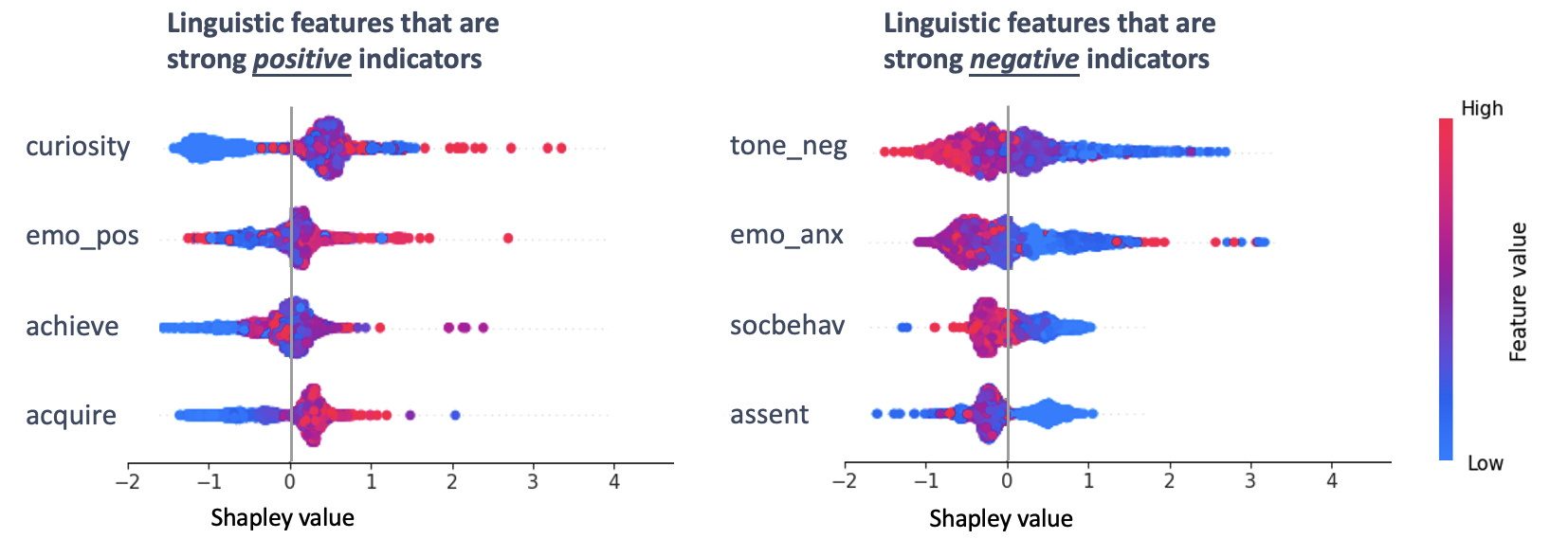}
    \caption{Fine-grained visualization of Shapley values with respect to those selected linguistic features across all possible feature combinations. The left hand side shows the four features that are strong positive indicators, while the right hand side shows the four features that are strong negative indicators. Points of red denote higher feature values, while points of blue represent lower values.}
    \label{fig:fine-shapley}
\end{figure}

The next step was to identify the most important linguistic features that can predict obtained peer acknowledgment.
Initially, we assessed the compatibility of five machine learning models — Random Forest~\cite{breiman2001random}, Lasso Regression (Lasso)~\cite{tibshirani1996regression}, K-Nearest Neighbors (KNN)~\cite{song2017efficient}, and ElasticNet~\cite{zou2005Elastic}, Gradient Boosting Regressor (GBR)~\cite{friedman2001GBR} — with our dataset, using R-squared (R$^2$) and Mean Squared Error (MSE) as the metrics for evaluating the predictive efficacy of these regression models. 
\new{The details of model training and overfitting prevention are described in Appendix~\ref{B}.}
As shown in Table~\ref{tab:result_table}, GBR emerged as the best model, exhibiting the highest R$^2$ and the lowest MSE values on both cross-validation and testing data.
Subsequently, we employed the Shapley value technique through the SHAP library\footnote{\href{https://shap.readthedocs.io/en/latest/}{https://shap.readthedocs.io/en/latest/}} with the GBR model ($M$), as defined in Eq.~\ref{eq:shapley}.
According to the output Shapley values in Figure~\ref{fig:Feature_importance}, most of the selected indicators in the affect (e.g., ``\textit{curiosity}'', ``\textit{emo\_pos}''), cognition (e.g., ``\textit{insight}'', ``\textit{tentat}''), and motivation (e.g., ``\textit{achievement}'', ``\textit{acquire}'') dimensions exhibit significant positive predictive effects on peer acknowledgement.


In contrast, among the twenty linguistic indicators, only four of them exhibit moderate to strong negative predictive value on peer acknowledgement. 
The negative indicators include ``\textit{tone\_neg}'' and ``\textit{emo\_anx}'' under the affect dimension, ``\textit{socbehav}'' and ``\textit{assent}'' under the social dimension. 
Two cognitive indicators, ``\textit{differ}'' (i.e., expressions of discrepancies) and ``\textit{certitude}'' (i.e., boasting of certainty), and one social indicator ``\textit{socref}'' (i.e., social reference), also show a negative effect on peer acknowledgement, but these effects are relatively small compared to others.

Figure~\ref{fig:fine-shapley} shows a more fine-grained view of the Shapley values across all feature combinations, specifically focusing on the top four positive indicators and the top four negative indicators. 
As illustrated, expression of positive affect and motivation are strong predictors of peer acknowledgement, while social indicators and negative emotions have a contradictory impact in prediction. 
The following section will discuss the implications in detail.
\section{Discussion}
\subsection{Peer Acknowledgement Facilitates Learners’ Social Annotation Behaviours} 
The primary focus of our first research question is to investigate the impact of peer acknowledgement on learners' annotation behaviour. Previous research has consistently shown that active participation in social annotation activities correlates with improved learning outcomes and a greater benefit from insights contributed by peers (e.g., ~\cite{lin2012participatory, johnson2010individual, sun2016online}). Furthermore, peer support plays a pivotal role in motivating individuals to engage in autonomous learning environments, where instructors typically take a less active role in guiding discussions ~\cite{feeley2012assessment, perrin2014features}. 


Given the asynchronous nature of online interaction in social annotation, our investigation focuses on the lagged effect of peer acknowledgement rather than the immediate correlation between received peer acknowledgement and social annotation contributions within the same time frame. We specifically explored how peer acknowledgement received in previous activities influences individuals' subsequent annotation behaviours, including initiating new posts and responding to others' contributions. Our findings provide strong support for the positive impact of peer support on learner engagement in social annotation. In particular, learners who received more peer acknowledgement in previous social annotation activities were more likely to contribute by initiating posts and responding to others. This increased engagement enriches the learning experience and interactions within the community.

Our findings align with the components of social presence outlined in the Community of Inquiry framework ~\cite{garrison2005facilitating,garrison2007researching}, suggesting that a sense of genuine connection with others enhances engagement in related activities ~\cite{rovai2007facilitating}. Additionally, our results are consistent with positive psychology ~\cite{lopez2009oxford, lopez2011encyclopedia} and teamwork ~\cite{salas2005there} theories, which propose that individuals who feel welcomed in a community, such as a social annotation group, tend to have a higher sense of self-value in their involvement ~\cite{reis2003toward}. This, in turn, fosters mutual trust with peers from the same community, motivating individuals to invest greater effort in contributing to the community.

\subsection{Influence of Psychological Features on Peer Acknowledgement in Written Text} 
Furthermore, we have identified specific psychological themes within the content of written annotations that hold particular significance in eliciting peer acknowledgment.
These findings shed light on the key psychological themes that significantly influence peer acknowledgment within digital social annotation environments. Notably, several dimensions of psychological factors emerged as strong predictors of peer acknowledgment.
\new{The practical implications for students, teachers, and other practitioners related to each identified theme are also discussed below.}

\textbf{Positive Affect.}  Our findings reveal that expressions of positive affect, such as curiosity, positive emotion, and a positive tone, exhibit a strong positive impact on peer acknowledgment. Learners who convey positive emotions and maintain an optimistic tone in their annotations are more likely to receive peer acknowledgement. This aligns with previous research highlighting the role of positivity in fostering social interactions and engagement ~\cite{youssef2013engagement, oishi2007residential}. 
Conversely, expressions of negative affect, including a negative tone or anxiety, could hinder peer acknowledgment. 
\new{The finding responds to the call for  theoretical buy-in regarding the perceived analytic value by examining the affective components of students’ online discussion \cite{wise2014learning}. Recognizing the impact of positive affect on peer acknowledgment, educators can foster a positive online learning environment. Incorporating elements that cultivate a sense of optimism and camaraderie within the digital community can contribute to a more supportive and collaborative learning space.}

\textbf{Openness for Learning and Discussion.} 
Among the cognitive indicators, annotations characterized by elements such as causation, tentative language, and insight tend to receive more peer acknowledgment. 
For instance, one sample annotation scored high in \textit{``causation''} in the cognitive process stating, \textit{``I would love to see more of this type of technology integrated into the classroom as it helps students learn, creates new friendships, and teaches them to cooperate''}.
By using the logistic conjunction and elaboration on the statement, the annotation is more likely to foster peer acknowledgment. 
However, expressions of discrepancies and certitude, which reflect a degree of bravado, exhibit a slightly adverse predictive direction.
For example, an annotation scored high in \textit{``certitude''} stated, \textit{``100\% I definitely agree''}, shows a high degree of certainty but it also presents less effort in the cognitive process and does not lead to further discussion. 
This suggests that boasting certainty or showing discrepancies in annotations may deter peer acknowledgment, while openness to discussion and providing evidence for critical thinking tend to attract more recognition.
\new{The theme offers insights for students; understanding that annotations emphasizing critical thinking and open discussion tend to attract more recognition provides them with a strategic approach. It also provides instructors with guidance on setting evaluation criteria. Encouraging cognitive engagement and discouraging excessive certainty or discrepancies in social annotation can contribute to more meaningful and acknowledged interactions among students.}

\textbf{Expression of Motivation.} Another influential indicator is the expression of motivation, particularly in terms of learners' desire to achieve, acquire knowledge, seek affiliation, and exercise power. Annotations reflecting high motivation levels are positively associated with peer acknowledgment. One sample annotation received high peer acknowledgement and also expressed high motivation in the learning context stating: \textit{``Three-dimensional dissection is really interesting. I took two anatomy courses last year where we missed the experience of dissecting bodies in real life. Hopefully, I get to experience both dissections in Virtual reality and in real life''}.
These words \textit{``hopefully''} and \textit{``get to''} demonstrate a sense of motivation in gaining more experience. This suggests that learners who demonstrate a strong drive and enthusiasm for learning are more likely to attract recognition from their peers. In essence, motivation plays a pivotal role in shaping the dynamics of peer acknowledgment within digital social annotation environments.
\new{This insight can motivate students to articulate their enthusiasm, fostering a more engaged and active community within the online learning platform.
}

\textbf{Social Aspects: Matter Less.} Additionally, the social theme identified from linguistic features does not contribute much to predicting peer acknowledgement, which is aligned with previous study ~\cite{wickersham2006content}. For example, words such as \textit{``okay''} and \textit{``yeah''}, typically associated with informal content like social conversations and casual dialogues, do not significantly contribute to prompting peer acknowledgment. This implies that learners' engagement in more formal and substantive discussions is more likely to elicit peer acknowledgement in our learning setting. 

To encapsulate, we identified three psychological themes that have relative importance in fostering peer acknowledgement, converting dimensions of affect, motivation, and cognition. Specifically, learners who convey positive affect, exhibit high motivation and engage in cognitive processes that encourage critical thinking tend to garner more peer acknowledgment. 
Conversely, expressions of negative affect and engagement in informal social behaviour are associated with lower peer acknowledgement. 

\new{The findings provide insights for learners, teachers, and technology developers by emphasizing the significance of affect, motivation, and cognitive engagement in shaping peer interactions. 
Teachers can foster the development of a supportive learning community by emphasizing these dimensions through the establishment of evaluation criteria. This encourages students to contribute to a positive and encouraging atmosphere.
The identified psychological themes also provide practitioners and technology developers with novel insights for designing effective feedback mechanisms by embedding a learning analytic approach. Building on previous studies that found significant results indicating the impact of learning analytics on learners’ online discussions compared to instructor supervision ~\cite{cerro2020impact}, technology developers can design feedback systems that acknowledge and reinforce positive contributions, thereby motivating students to actively participate and contribute meaningfully to discussions. For example, automatic detection algorithms can be implemented to filter students’ annotations with the desired themes and provide supportive prompts, such as awards and hints, to those lacking in addressing these themes, along with encouraging reactions.}

\subsection{Limitation and Future Direction} 
Despite our findings, our study has several limitations. 
First, the topics selected for social annotation relied on social sciences within our learning context. 
Therefore, the generalizability of our findings to other subject-specific contexts may be limited. 
Future studies could explore participants' learning trajectories in other domains such as engineering or mathematics, which may involve different content designs and workloads for assigned social annotation tasks.

\new{Though valuable for analyzing written language, LIWC has some limitation compared to advanced NLP approaches. For example, its pre-defined dictionary may not cover emerging terms, and it may struggle to capture complex sentences. Still, considering its wide application in psychology ~\cite{huangemotion}, sociology ~\cite{zhengthinking}, and computer science ~\cite{lidetection} to analyze and understand language content and the efficiency of the analytic process, it is appropriate to use this tool in our social annotation context.}
To delve deeper into this context, a more profound comprehension of the psychological factors that facilitate peer interaction could be attained by integrating qualitative analysis into future study design. For instance, future research endeavours could encompass human coding of the extracted written texts and the analysis of individuals' reflective diaries or interviews, delving into their experiences with social annotation activities.
\section{Conclusion}
Our research contributes to a deeper understanding of the role of peer acknowledgement in promoting active participation and collaboration among learners in computer-supported collaborative learning environments. 
Drawing upon established theoretical frameworks and empirical evidence, we shed light on the underlying mechanisms that drive increased engagement and contribution within these learning communities. 
Our study underscores the significance of peer acknowledgement and its positive impact on individuals' learning behaviors, while also highlighting the intricate interplay of psychological factors that shape peer acknowledgment dynamics within digital social annotation environments. 
The findings provide empirical evidence for practitioners in education and training using social annotation to promote peer acknowledgement through encouraging positive affect, expression of motivation, and openness for learning and discussion.
Furthermore, educational technology developers and programmers can utilize learners' linguistic features to design computer-human interactions that provide adaptive hints to encourage effective interactions in digital social annotations.


\bibliographystyle{ACM-Reference-Format}
\bibliography{references}

\appendix

\section{Explanation of Mixed-Effects Linear Regression Model}
\label{A}
\new{In our study, we integrated a linear mixed-effects model in the cross-lag regression analysis to investigate the influence of peer acknowledgment on learners’ social annotation behaviors. The model accounts for individual-specific effects, which is crucial in our context where the effect of peer acknowledgment may vary across participants: $Y_{ij} = \beta_0 + \beta_1\cdot X_{ij} + u_{1j}\cdot X_{ij} + \epsilon_{ij}$.}
\begin{itemize}[leftmargin=*]
    \item \( Y_{ij} \): Dependent variable representing the social annotation behavior (either the number of annotations initiated or responses given) for the \( i \)-th observation of the \( j \)-th participant.
    \item \( X_{ij} \): Independent variable indicating peer acknowledgment for the \( i \)-th observation of the \( j \)-th participant.
    \item \( \beta_0 \): Fixed intercept of the model.
    \item \( \beta_1 \): Fixed slope, representing the overall effect of peer acknowledgment on social annotation behavior.
    \item \( u_{1j} \): Random slope for the predictor \( X \), specific to each participant \( j \). This component allows the model to capture individual differences in the relationship between peer acknowledgment and social annotation behaviors.
    \item \( \epsilon_{ij} \): Residual error for the \( i \)-th observation of the \( j \)-th participant.
\end{itemize}

\new{In summary, the model incorporates a random slope for the grouping variable (participant ID), denoted as $u_{1j}$ in the formula.
This approach acknowledges that the effect of peer acknowledgment on annotation behaviors may not be the same for all participants, thus providing a more nuanced understanding of the underlying relationships.
This mixed-effects linear regression model with a random slope component is pivotal in our analysis, as it aptly handles the variability across individual participants, enhancing the robustness and interpretability of our findings.}
\vspace{-2mm}
\section{Model Training Details}
\label{B}
\vspace{-6mm}
\new{\paragraph{Training/Test Split} The dataset, consisting of 1989 data entries, was divided into a training set (70\%) and a testing set (30\%). This division resulted in approximately 1392 data entries for training and about 597 entries for testing.}
\vspace{-2mm}
\new{\paragraph{Cross-Validation} We employed $k$-fold cross-validation with $k$ set to 5 on the training data. This method divides the training data into five subsets. The model is trained on four subsets and validated on the remaining one, with this process being repeated five times, each with a different subset as validation data.}
\vspace{-2mm}
\new{\paragraph{Overfitting Prevention} Several strategies were implemented to mitigate overfitting. These included applying regularization techniques in models such as Lasso and ElasticNet, and limiting the complexity of the Random Forest and Gradient Boosting Regressor (GBR) models by controlling the maximum depth of trees. Performance metrics were consistently monitored across both cross-validation and testing phases.}
\vspace{-2mm}
\new{\paragraph{Comparison of Cross-Validation and Testing Results} Overfitting is often indicated when a model performs significantly better on training data compared to unseen test data. In our case, the cross-validation and testing results for each model are quite close. For instance, the GBR, our best model, shows R-squared values of 0.9231 in cross-validation and 0.9123 in testing. This closeness in performance metrics across both phases suggests good generalization and is a positive sign against overfitting.}

\end{document}